\documentclass[a4paper,10pt,aps,amsfonts,amsmath,nofootinbib,showpacs,byrevtex,notitlepage, prd
]{revtex4-1}

\usepackage{graphicx}
\usepackage[caption=false]{subfig}
\usepackage{bm,bbm}

\newcommand{\cN}{{\cal N}}
\newcommand{\cD}{{\cal D}}

\newcommand{\pa}{\parallel}
\newcommand{\sphere}{{\mathrm S}}
\newcommand{\RR}{\mathbbm{R}}

\newcommand{\be}{\begin{equation}}
\newcommand{\ee}{\end{equation}}

\DeclareMathOperator{\Det}{Det}
\DeclareMathOperator{\Tr}{Tr}

\newcommand{\lk}{\left(}
\newcommand{\rk}{\right)}

\newcommand{\im}{\mathrm{i}}
\newcommand{\e}{\mathrm{e}}

\renewcommand*{\vec}[1]{\bm{\mathrm{#1}}}

\newcommand{\vx}{\vec{x}}
\newcommand{\ve}{\vec{e}}
\newcommand{\vp}{\vec{p}}
\newcommand{\vA}{\vec{A}}
\newcommand{\vB}{\vec{B}}
\newcommand{\vD}{\vec{D}}
\newcommand{\vy}{\vec{y}}
\newcommand{\vq}{\vec{q}}

\providecommand*{\coloneq}{\mathrel{\mathop:}=}

\renewcommand*{\d}[1][]{\mathop{\mathrm{d}^{#1}}\mkern-4mu}
\newcommand*{\ddbar}[1][]{\mathop{\mathrm{d}\mkern-7mu\mathchar'26\mkern-1mu^{#1}}\mkern-4mu}

\begin{document}

\title{Finite-temperature Yang-Mills theory in the Hamiltonian approach in Coulomb gauge from a compactified spatial dimension}

\author{J. Heffner and H. Reinhardt }
\affiliation{Institut f\"ur Theoretische Physik, Eberhard-Karls-Universit\"at T\"ubingen,
Auf der Morgenstelle 14, 72076 T\"ubingen, Germany}
\pacs{
03.70.+k, 
11.15.-q, 
12.38.Aw, 
}
\keywords{Hamiltonian approach, Coulomb gauge, Finite temperature}
\date{\today}

\begin{abstract}
Yang-Mills theory is studied at finite temperature within the Hamiltonian approach 
in Coulomb gauge by means of the variational principle using a Gaussian type ansatz for the vacuum wave functional. 
Temperature is introduced by compactifying one spatial dimension. As a consequence the finite temperature behavior is encoded in the vacuum wave functional calculated on the 
spatial manifold $\RR^2 \times \sphere^1 (L)$ where $L^{-1}$ is the temperature. The finite-temperature equations of motion are obtained by minimizing the vacuum energy density to two-loop order. We show analytically 
that these equations yield the correct zero-temperature limit while at infinite temperature they reduce to the equations of the $2$+$1$-dimensional theory in 
accordance with dimensional reduction. The resulting propagators 
are compared to those obtained from the grand canonical ensemble where an additional ansatz for the density
matrix is required.
\end{abstract}

\maketitle

\section{Introduction}

In recent years substantial efforts have been undertaken to study QCD non-perturbatively in the continuum. The  main motivation is, of course, to 
understand the low-energy properties of QCD such as confinement and spontaneous breaking of chiral symmetry. Another motivation for these studies
is the understanding of the phase diagram of QCD since the lattice approach can not deal with large chemical potentials due to the notorious
sign  problem.

In Refs.~\cite{Feuchter:2004mk,Feuchter:2004gb,Reinhardt:2004mm} 
a variational approach to the Hamilton formulation of QCD in Coulomb gauge was 
developed.\footnote{The relation of the present
approach \cite{Feuchter:2004mk} to previous variational calculations \cite{Schutte:1985sd,Szczepaniak:2001rg} 
in Coulomb gauge is discussed in detail in Ref.~\cite{Greensite:2011pj}. Furthermore an analogous variational approach was recently developed in a covariant formulation in Landau gauge \cite{Quandt:2013wna}.}
In the Yang-Mills sector this approach used initially  Gaussian type variational ans\"atze for the vacuum wave functional.
With the energy density calculated up to two-loops this variational approach has given a decent description of the infrared properties of the various propagators. Among
other results one has found an infrared diverging gluon energy \cite{Feuchter:2004mk,Epple:2006hv}
 and a perimeter law for the spatial 't Hooft loop \cite{Reinhardt:2007wh} as well as a linear rising static 
color potential \cite{Epple:2006hv}. 
These are all signatures of confinement. Furthermore, it was shown that the inverse ghost form factor
of the Hamiltonian approach in Coulomb gauge represents the color dielectric function of the Yang-Mills vacuum 
\cite{Reinhardt:2008ek}. 
By the so-called horizon condition \cite{Zwanziger:1998ez} this quantity vanishes in the infrared, which implies that the Yang-Mills vacuum behaves as a dual superconductor.

The variational Hamiltonian approach in Coulomb gauge was extended in Refs.~\cite{Reinhardt:2011hq,Heffner:2012sx} 
to finite temperatures by assuming a quasi-particle
ansatz for the density operator of the grand canonical ensemble and minimizing the free energy.
The obtained infrared exponents of the gluon and ghost 
propagators show a drastic change at the deconfinement phase transition from which a critical temperature of about $T_C \approx 275 \dots 290$ MeV was obtained
for SU$(2)$ \cite{Heffner:2012sx}. In Ref.~\cite{Reinhardt:2012qe,Reinhardt:2013iia} 
the effective potential of the Polyakov loop was calculated within the Hamiltonian approach in Coulomb  gauge. Thereby the 
temperature was introduced by compactifying one spatial dimension. 
Introducing the temperature into the Hamiltonian approach by compactifying a spatial dimension has the advantage that no ansatz for the density matrix (of the grand canonical ensemble) is required.
The finite-temperature theory is fully encoded in the ground state on the spatial manifold $\RR^2 \times \sphere^1 (L)$ with $L^{-1}$ being the temperature, see Refs.~\cite{Reinhardt:2012qe,Reinhardt:2013iia} for more details.
Furthermore only this way of introducing the temperature makes the Polyakov loop accessible to the Hamiltonian approach, which uses the Weyl gauge $A_0=0$.
From the effective Potential of the Polyakov loop a critical temperature similar to that of the grand canonical ensemble is extracted.
Furthermore, the correct order of the phase transition is obtained for both color groups SU($2$) and SU($3$).
However, in the actual calculation of the Polyakov loop potential the propagators from the zero-temperature variational solutions were used, while, in principle, the finite-temperature self-consistent solutions should be used instead.
Therefore in the present paper we extend the Hamiltonian approach to Yang-Mills theory in Coulomb gauge developed in Ref.~\cite{Feuchter:2004mk} to the spatial manifold $\RR^2 \times \sphere^1 (L)$. 

The rest of the paper is organized as follows: In the next section we derive the basic equations of a variational treatment of Yang-Mills theory in Coulomb gauge on
to the spatial manifold $\RR^2 \times \sphere^1 (L)$.
Due to the explicit breaking of O$(3)$ symmetry on this spatial manifold this will require to generalize the ansatz for the vacuum wave functional of Ref.~\cite{Feuchter:2004mk} 
distinguishing between the components of three vectors parallel and orthogonal to the compactified spatial dimension. In Sect.~\ref{sectionIII} 
we investigate the low- and high-temperature limits of this approach. The infrared analysis and the renormalization of our equations of motion are carried out in Sect.~\ref{sec:IRanRen}. In Sect.~\ref{sect:numerical} we present our numerical results for the gluon
and ghost propagators. Some concluding remarks are given in Sect.~\ref{sect:concl}.

\section{Finite-temperature Yang-Mills theory from a compactified  spatial dimension}\label{IV}

Following Ref.~\cite{Reinhardt:2013iia} we introduce the temperature into the Hamiltonian approach to a quantum field theory by compactifying
one spatial dimension, for which we choose the 3-axis. The partition function is then given
\be
\label{190-7-1}
Z (L) = \lim\limits_{l \to \infty} \Tr \e^{- l E_0 (L)} \, ,
\ee
where $E_0 (L)$ is the energy of the vacuum state on the spatial manifold $\RR^2 \times \sphere^1 (L)$ and $l \to \infty$ is the length of the 
uncompactified spatial dimensions. Note that in the present approach the whole temperature dependence is contained in the ground state energy $E_0 (L)$ and 
is completely encoded in the vacuum state. Excited states would only contribute when $l$ is kept finite, which corresponds to restricting one of the true spatial dimension
to a finite length, as one encounters in the Casimir effect for two parallel infinite plates separated by a distance $l$. 

\subsection{Yang-Mills theory in Coulomb gauge on $\RR^2 \times \sphere^1 (L)$}

On $\RR^2 \times \sphere^1(L)$ the gauge field satisfies periodic boundary conditions in the third (compactified) dimension 
\be
\label{65}
\vA (\vx_\perp, L / 2) = \vA (\vx_\perp, - L / 2) \, .
\ee
Here and in the following we denote by $\vx_\perp$ the component of $\vx$ orthogonal to the compactified dimension.
The Yang-Mills Hamiltonian in Coulomb gauge \cite{Christ:1980ku}
\be
\label{59}
\vec{\partial} \vA = \vec{\partial}_\perp \vA_\perp + \partial_3 A_3 = 0
\ee
on $\RR^2 \times \sphere^1(L)$ is formally the same as on $\RR^3$
\be
\label{60}
H  = \frac{1}{2} \int_L \d^3 x  \lk g^2 J_A^{- 1} \vec{\Pi} J_A \vec{\Pi} + 
\frac{1}{g^2} \vB ^2 \rk + H_\mathrm{C}  \, ,
\ee
except that the integration measure over the spatial manifold is defined by
\be
\label{217-2}
\int_L \d^3 x := \int \d^2 x_\perp \int^{L/2}_{-L/2} \d x_3 \, .
\ee
In Eq.~(\ref{60}) $\Pi^a_k (x) = - \im \delta / \delta A^a_k (x)$ is the momentum operator, $\vB$ is the non-Abelian magnetic field and 
\be
\label{61}
J_A = \Det (- \hat{\vD} \vec{\partial})
\ee
is the Faddeev-Popov determinant, where $\hat{D}^{ab}_k = \delta^{ab} \partial_k + f^{acb} A^c_k$ is the covariant derivative in the 
adjoint representation of the color group with $f^{abc}$ being the structure constants. Finally,
\be
\label{62}
H_\mathrm{C} = \frac{g^2}{2} \int_L \d^3 x \int_{L} \d^3 y \, J_A^{- 1} \rho^a (\vx) J_A 
F^{ab} (\vx, \vy) \rho^b (\vy)
\ee
is the Coulomb term, which arises from the elimination of the longitudinal components of the momentum operator by means of 
Gauss' law. Here 
\be
\label{63}
F^{ab} (\vx, \vy) = \langle \vx | \left[ (- \hat{\vD} \vec{\partial})^{- 1} (- \vec{\partial}^2) (- \hat{\vD} \vec{\partial})^{- 1}
\right]^{ab} | \vy \rangle 
\ee
is the so-called Coulomb kernel, which is the non-Abelian generalization of the ordinary Coulomb law. 

To find the vacuum state we use the variational principle with the trial vacuum wave functional \cite{Feuchter:2004mk} 
\be
\label{66}
\psi[A] = \cN J_A^{- 1/2} \exp \lk - \frac{1}{2 g^2} \int_L \d^3 x \int_L \d^3 y \, A^a_k (\vx) \omega_{kl} (\vx, \vy) A^a_l (\vy) \right) \, .
\ee
This ansatz preserves global color symmetry. The pre-exponential factor was chosen to cancel the Faddeev-Popov determinant in the functional
integral for the expectation values 
\be
\label{67}
\langle \ldots \rangle = \int \cD A \,J_A \psi^*[A] \ldots \psi[A] \, .
\ee
Due to the Coulomb gauge condition the variational kernel $\omega_{kl} (\vx, \vy)$ can be chosen transversal. On the spatial manifold $\RR^3$, due to O($3$) invariance, transversality implies 
\be
\label{68}
\omega_{kl} (\vx, \vy) = t_{kl} (\vx) \omega (\vx, \vy) \, ,
\ee
where
\be
t_{kl} (\vx) = \delta_{kl} - \frac{\partial^x_k \partial^x_l}{\Delta_x}
\ee
is the usual transversal projector in $\RR^3$. However, on $\RR^2 \times \sphere^1(L)$ the O($3$) symmetry is broken to O($2$) and the ansatz
(\ref{68}) has to be generalized to 
\be
\label{70}
\omega_{kl} (\vx, \vy) = t^\perp_{kl} (\vx) \omega_\perp (\vx, \vy) + t^{\pa}_{kl} (\vx) \omega_{\pa} (\vx, \vy) \, .
\ee
Here
\be
\label{71}
t^\perp_{kl} (\vx) = (1 - \delta_{k3}) \lk \delta_{kl} - \frac{\partial^x_k \partial^x_l}{\Delta^\perp} \right) (1 - \delta_{l3}) 
\ee
is the transversal projector in the subspace $\RR^2$ orthogonal to the compactified  3-axis with $\Delta_\perp = \partial^2_1 + \partial^2_2$ and
furthermore
\be
\label{72}
t^{\pa}_{kl} (\vx) = t_{kl} (\vx) - t^\perp_{kl} (\vx)
\ee
is the corresponding orthogonal complement. There are now two independent variational kernels, $\omega_{\pa}$ and $\omega_\perp$, corresponding to the compactified
dimension and the orthogonal subspace, respectively. To simplify the notation we introduce an index $Q \in \{ \pa, \perp\}$ to distinguish the compactified dimension 
from the space orthogonal to it.

The  projectors introduced above have the following properties 
\begin{align}
\label{73}
t^Q_{kl} (\vx) t^Q_{lm} (\vx) &= t^Q_{km} (\vx), \quad t^Q_{kl} (\vx) t_{lm} (\vx) = t^Q_{km} (\vx) \, , \nonumber\\
t_{kk} (\vx) & = 2  , \quad t^Q_{kk} = 1 \, .
\end{align}
With the trial ansatz (\ref{66}), (\ref{70}) the static gluon propagator is given by
\be
\label{74}
D^{ab}_{kl} (\vx, \vy) = \langle A^a_k (\vx) A^b_l (\vy) \rangle = \delta^{ab} \lk D^\perp_{kl} (\vx, \vy) + D^{\pa}_{kl} (\vx, \vy) \rk
\, 
\ee
where
\be
\label{75}
D^Q_{kl} (\vx, \vy) = \frac{1}{2} t^Q_{kl} (\vx) \omega^{- 1}_Q (\vx, \vy) \, .
\ee
 By translational invariance the kernels $\omega_Q (\vx, \vy)$ depend only on the relative coordinate 
$\vx - \vy$ and it is convenient to work in momentum space, which is defined here by
\be
\label{860-13c}
\omega (\vx) = \int_L \ddbar^3 p \, \e^{\im \vp  \vx} \omega (\vp) \coloneq \int \ddbar^2 p_\perp \frac{1}{L} \sum^\infty_{n = - \infty} \e^{\im \vp_\perp \vx_\perp} \e^{\im \omega_n x_3} \omega (\vp_\perp, \omega_n) \, ,
\ee
where $\ddbar^2 p_\perp = \d^2 p_\perp / (2 \pi)^2$ is the integration measure in the momentum plane orthogonal to the $3$-axis
and 
$\omega_n = 2 \pi n / L$
is the Matsubara frequency. From the  
representation (\ref{75}) it is seen that $\omega_Q (\vp_\perp, \omega_n)$ with 
$Q = \pa$ and $\perp$, respectively are the 
energies of the transverse gluon modes parallel and orthogonal to the compactified spatial dimension.  

\subsection{Equations of motion}

\begin{figure}
\centering
\includegraphics[width=.3\linewidth]{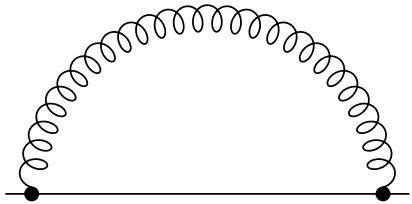}
\caption{Ghost self-energy in rainbow-ladder approximation. The full and wavy line represent, respectively, the ghost and gluon propagator.}
\label{fig1}
\end{figure}

The variational approach to Yang-Mills theory in Coulomb gauge on $\RR^2 \times \sphere^1 (L)$ can be worked out in the same way as it was done in Ref.~\cite{Feuchter:2004mk} on $\RR^3$ 
with the exception that the transverse degrees of freedom in $\RR^3$ are divided now into the degrees parallel and orthogonal to the 
compactified dimension, see Eq.~(\ref{72}). Accordingly there are two independent variational kernels $\omega_{\pa}$ and $\omega_\perp$.
In Ref.~\cite{Heffner:2012sx} the Coulomb term $H_\mathrm{C}$ (\ref{62}) was shown to be irrelevant in the gluon sector and will therefore be ignored below. 
Following Ref.~\cite{Feuchter:2004mk}, but using the more general ansatz (\ref{70}) for the vacuum wave functional, the variation of the energy, 
calculated up to two-loop order, results in the following two gap equations
\be
\label{81}
\omega^2_Q (\vp) = \vp^2 + \chi^2_Q (\vp) \, , \quad \quad Q \in \{ \pa, \perp\} \, ,
\ee
where $\vp^2 = \vp^2_\perp + \omega_n^2$ is the square of the $3$-momentum 
\be
\vp  = \vp_\perp + \omega_n \ve_3, \quad \omega_n = 2 \pi n / L
\ee
and
\be
\label{82}
\chi_Q (\vp) = \frac{g^2}{2} N_c \, t^Q_{kl}(\vp) \int_L \ddbar^3 q \, q_k q_l \, G (\vp + \vq) G (\vq).
\ee
is the ghost loop, referred to in the present context as ``curvature'' \cite{Reinhardt:2004mm}.
Here $G (\vp)$ is the ghost propagator defined by
\be
\label{76}
G  = \langle (- \hat{\vD} \vec{\partial} )^{- 1} \rangle.
\ee
In momentum space the ghost propagator is conveniently expressed by its form factor $d (\vp)$ defined by 
\be
\label{78}
G (\vp) = \frac{d (\vp)}{g \vp^2} .
\ee
The ghost form factor $d (\vp)$ expresses the deviation of QCD from QED and its inverse can be interpreted as the dielectric function of the Yang-Mills vacuum \cite{Reinhardt:2008ek}.
Calculating the ghost propagator with the vacuum wave functional (\ref{66})  in the rainbow-ladder approximation results in the following Dyson-Schwinger equation (DSE) for the ghost form factor 
\be
\label{79}
d^{- 1} (\vp) = \frac{1}{g} - \lk I^\pa_d (\vp) + I^\perp_d  (\vp) \right) \,  ,
\ee
where the loop integrals
\be
\label{80}
I^Q_d (\vp) =  N_c \int_L \ddbar^3 q \,\frac{p_i t^Q_{i j} (\vq) p_j}{\vp^2 }\frac{d (\vp + \vq)}{(\vp + \vq)^2} \frac{1}{2 \omega_Q (\vp) } 
\ee
arise from the ghost self-energy diagram, shown in fig.~\ref{fig1}. 
DSEs, in general, are functional differential equations and their solutions are specified only after providing appropriate boundary
conditions. For the ghost DSE we require (as on $\RR^3$) the horizon condition, which on $\RR^2 \times \sphere^1 (L)$ reads
\be
\label{eq:horizon}
d^{- 1} (\vp_\perp = 0, \omega_n = 0) = 0 \, .
\ee

The two gap equations (\ref{81}) and the ghost DSE (\ref{79}) are coupled through the loop terms 
and have to be solved simultaneously. In the 
IR regime this can be done analytically following the by now standard IR analysis \cite{Lerche:2002ep,Schleifenbaum:2006bq}, see Sect.~\ref{sectionIII}.
The numerical solution for these equations in the whole momentum regime will be given in Sect.~\ref{sect:numerical}.

\section{Zero- and high-temperature limits}\label{sectionIII}

Before we present the numerical results let us investigate the above obtained equations of motion in the high and low temperature limit. 

\subsection{The high-temperature limit}

In the high-temperature limit $T = L ^{-1} \to \infty$ the Matsubara frequencies $\omega_n = 2 \pi n T$ with $n \neq 0$ become infinite. Since 
for $p \to \infty$ all propagators vanish we need to keep in this limit in the above equations of motion only the lowest Matsubara frequency, $p_3 \equiv \omega_n  = 0$, implying the replacement (cf. Eq.~(\ref{860-13c}))
\be
\label{83}
\int_L \ddbar^3 p \to T \int \ddbar^2 p_\perp \, .
\ee
For $p_3 = 0$ we have 
\be
\label{84}
\lk p_k t^{\pa}_{kl} (q) p_l \right)_{p_3 = q_3 = 0} = 0 \, ,
\ee
so that the longitudinal part of the ghost self-energy vanishes,  $I^\pa_d (\vp_\perp,\omega_n=0) = 0$, and the ghost DSE (\ref{79}) reduces to 
\be
\label{85}
d^{- 1} (\vp_\perp, \omega_n  = 0) = \frac{1}{g} - I^\perp_d (\vp_\perp,  \omega_n   = 0) \, 
\ee
with 
\be
\label{86}
I^\perp_d (\vp_\perp, \omega_n  = 0) =  N_c T \int \ddbar^2 q_\perp \frac{p^\perp_i t_{ij} (\vq_\perp) p^\perp_j}{\vp_\perp^2} \frac{d (\vp_\perp 
+ \vq_\perp, 0)}{(\vp_\perp + \vq_\perp)^2} \frac{1}{2\omega_\perp (\vq_\perp, 0)} \, .
\ee
This is precisely the ghost DSE  in $d = 2$ spatial dimensions \cite{Feuchter:2007mq} 
provided we identify $\omega_\perp (\vq_\perp, 0)$ with the gluon energy and
\be
\label{87}
g^{(2)} = g \sqrt{T} \, , \quad \quad d^{(2)} (\vp_\perp) = d (\vp_\perp, 0) \sqrt{T}
\ee
with the coupling constant and ghost form factor, respectively, in $d = 2$. As is well known the dimension of the Yang-Mills coupling constant $g$ depends on the number of space-time dimensions.
In $d = 3$ spatial dimensions $g$ is dimensionless, while it has dimension $\sqrt{\mbox{energy}}$ in $d = 2$. Yang-Mills theory in  $d = 3$ is scale free but the temperature does introduce a 
scale. Thus, a relation like Eq.~(\ref{87}) should have been expected. From  Eq.~(\ref{84}) also follows that the ``longitudinal'' component of the 
curvature (\ref{82}) vanishes in the high-temperature limit, $\chi_{\pa}(\vp_\perp, \omega_n = 0) = 0$, and accordingly the longitudinal gap equation (\ref{81}) reduces to 
\be
\label{88}
\omega^2_\pa (\vp_\perp, \omega_n = 0) = \vp^2_\perp \, 
\ee
implying that the gluon degree of freedom parallel to the compactified dimension becomes free at $T \to \infty$.

With Eq.~(\ref{83}) we find for the transversal curvature (\ref{82}) 
\be
\label{89}
\chi_\perp (\vp_\perp, \omega_n = 0) = \frac{g^2 T N_c}{2} \int \ddbar^2 q_\perp \, \frac{q^\perp_i t_{ij} (\vp^\perp) q^\perp_j }{ \vq_\perp^2}
\frac{d (\vp_\perp + \vq_\perp, 0) d (\vq_\perp, 0)}{g^2 (\vp_\perp + \vq_\perp)^2}
\, ,
\ee
which, with the identification (\ref{87}), becomes the curvature in $d = 2$ spatial dimensions \cite{Feuchter:2007mq}. 
Accordingly the transversal gap equation (\ref{82}) becomes that of two dimensions
\be
\label{90}
\omega^2_\perp (\vp_\perp, \omega_n=0) = \vp^2_\perp + \chi^2_\perp (\vp, \omega_n=0) \, .
\ee
Thus for $T \to \infty$ we find complete dimensional reduction: Yang-Mills theory in $3$+$1$ dimension reduces in the high-temperature limit to the $2$+$1$ dimensional theory 
with 
an additional ``Higgs field''  given by the temporal component of  the gauge field, $A_0$. In the present Hamiltonian approach this Higgs field does not show up  since we work here in the 
Weyl gauge  $A_0 = 0$. 

\subsection{The zero-temperature limit}

As the temperature decreases more and more Matsubara frequencies have to be included. In the limit $T \to 0$
all Matsubara frequencies contribute and one has to carry out the whole infinite sum. For small $T$ it is advantageous to carry out the summation over the Matsubara frequency by means of 
the relation \cite{Reinhardt:2013iia}
\be
\label{469-12-X1}
\frac{1}{L} \sum^\infty_{n = - \infty} f (\omega_n) = \frac{1}{2 \pi} \int^\infty_{- \infty} \d z\,f (z) \sum^\infty_{k = - \infty} \e^{\im k L z} \, ,
\ee
which follows
 from the Poisson resummation formula.\footnote{When Eq.~(\ref{469-12-X1}) is applied to functions $f (\omega_n)$ known only 
numerically for the discrete Matsubara frequencies $\omega_n$ it is understood that 
the function $f (z)$ arises from $f (\omega_n)$ by smoothly interpolating the latter in the intervals between neighboring Matsubara
frequencies. In fact, for $L \to \infty$, where the Poisson resummation (\ref{469-12-X1}) is convenient, the spacing
between neighboring Matsubara frequencies tends to zero.} 

In order to carry out the zero-temperature limit in the equations of motion given in section~\ref{IV} consider a generic loop integral, which by means of Eq.~(\ref{469-12-X1}) can be written as 
\be
\label{92}
\frac{1}{L} \sum^\infty_{n = - \infty} \int \ddbar^2 p_\perp f (\vp_\perp, \omega_n) = \int \ddbar^3 p f (\vp_\perp, p_3)
\sum^\infty_{k = - \infty} \e^{\im k L p_3} \, .
\ee
Here we have put $z = p_3$ and $\ddbar^3 p =\d^3 p /(2 \pi)^3$ denotes the usual integration measure in $3$-dimensional momentum space. 
As $L \to \infty$ the O($3$) invariance is restored and all propagators and correlation functions
 will depend only on the 
modulus of the 3-momentum, i.e. we have
\be
\label{1174-23}
f (\vp_\perp, p_3) = f \lk \sqrt{\vp^2_\perp + p^2_3} \right) \equiv f (p) \, .
\ee
It is then convenient to use spherical coordinates in momentum space $(p, \theta, \varphi)$. Carrying out the 
integration over $\theta$ we obtain from Eq.~(\ref{92})
\be
\label{1180-23-1}
\frac{1}{L} \sum^\infty_{n = - \infty} \int \ddbar^2 p_\perp f \lk \sqrt{p^2_\perp + \omega^2_n} \right) = 
\int \ddbar^3 p f (p) \sum^\infty_{k =  - \infty} \frac{\sin (k L p)}{k L p} \, .
\ee
Obviously the term $k = 0$ reproduces the $T =0 $ result. The remaining terms vanish for $L \to \infty$.
With the relation (\ref{1180-23-1}) all equations given above reduce for $L \to \infty$ 
to the zero-temperature equations derived in Ref.~\cite{Feuchter:2004mk}. 

\section{Infrared analysis and renormalization}
\label{sec:IRanRen}
In Ref.~\cite{Schleifenbaum:2006bq} an IR-analysis of the zero-temperature equations of motion was carried out using the power-law ans\"atze (for $p \to 0$)
\be
\label{512-13}
\omega (p) \sim p^{- \alpha} \, , \quad \quad d (p) \sim p^{- \beta} \, .
\ee
Assuming a bare ghost-gluon vertex and the horizon condition $d^{- 1} (p = 0) = 0$, i.e. $\beta > 0$, one obtains from the ghost DSE the sum rule 
\be
\label{517-13-2}
\alpha = 2 \beta + 2 - d \, ,
\ee
where $d$ is the number of spatial dimensions. From the gap equation one finds then two solutions for $d = 3$ 
\be
\label{522-13-3}
\beta= 1  \, , \quad \quad \beta \approx 0.7952
\ee
and a single solution for $d = 2$
\be
\label{527-13-4}
\beta = 0.4\, .
\ee
The IR exponents extracted from the numerical $(T = 0)$ solutions obtained 
in Refs.~\cite{Feuchter:2004mk,Epple:2006hv} 
(see also Ref.~\cite{Heffner:2012sx} and \cite{Feuchter:2007mq}) are in excellent agreement with the analytic results of this infrared analysis.
From the result of the previous section it is clear that the propagators on $\RR^2 \times \sphere^1 (L)$ with the Matsubara frequency $n = 0$ will exhibit the $d = 3$ IR exponents (\ref{522-13-3}) in the low temperature regime and the $d = 2$ IR exponents (\ref{527-13-4}) in the high-temperature regime.
In this context it is worth mentioning that in the finite-temperature extension of the Hamiltonian approach to Yang-Mills theory in Coulomb gauge within the usual canonical ensemble given in Ref.~\cite{Heffner:2012sx} the IR-analysis can be carried out  in the high-temperature limit yielding an IR exponent of $\beta =0.5$.
In the corresponding numerical studies the deconfinement phase transition was accompanied by a rapid change of the IR exponents from the $(T = 0, d = 3)$-values (\ref{522-13-3}) to the high-temperature value of $\beta =0.5$, close to the $2$-dimensional value (\ref{527-13-4}).

As shown in the previous section for $L\to \infty$ ($T\to 0$)  the equations of motion of the present approach reduce indeed to the zero-temperature equations of Ref.~\cite{Feuchter:2004mk}.
By compactifying one spatial dimension no (additional) UV singularity can be introduced.
Furthermore, for large momenta $p \gg T$ the temperature becomes irrelevant.
Thus our equations of motion on $\RR^2 \times \sphere^1 (L)$  will have the same UV behavior as on $\RR^3$ and can be renormalized as in the zero-temperature case.
For $T=0$ the ghost loop $\chi(p)$ (\ref{82}) is linearly UV divergent, while the loop integrals $I^Q_d(p)$ (\ref{80}) in the ghost DSE are logarithmically divergent.
To renormalize the ghost DSE (\ref{79}) we use Eq.~(\ref{72}) to rewrite it in the form
\begin{multline}
\label{eq:ghostren}
d^{- 1} (\vp) = \frac{1}{g} - N_c \int_L \ddbar^3 q \, \left[1 -(\hat{\vp} \hat{\vq})^2\right] \frac{d (\vp + \vq)}{(\vp + \vq)^2} \frac{1}{2 \omega_\perp (\vq) } \\
+  N_c \int_L \ddbar^3 q \,\frac{p_i t^\pa_{i j} (\vq) p_j}{\vp^2 }\frac{d (\vp + \vq)}{(\vp + \vq)^2} \lk \frac{1}{2 \omega_\pa (\vq) } -\frac{1}{2 \omega_\perp (\vq) }   \rk.
\end{multline}
To renormalize the ghost DSE it would, in principle, be sufficient to subtract from the first loop integral Eq.~(\ref{eq:ghostren}) its zero-temperature counterpart at a renormalization scale $\mu_d$.
Although being correct this procedure is not feasible in an iterative solution of the ghost DSE.
Therefore we renormalize the ghost DSE by subtracting the first (temperature-dependent) loop integral at a scale $\mu_d$, which is chosen in the UV such that the horizon condition (\ref{eq:horizon}) is fulfilled.\footnote{This introduces de facto a temperature-dependent renormalization constant. The same renormalization procedure was used in the study of the grand canonical ensemble in Ref.~\cite{Heffner:2012sx}.}
Analogously we renormalize the curvature $\chi^Q(\vp)$ by subtracting it at a scale $\mu_\chi$
\be
\chi^Q(\vp) \longrightarrow \chi^Q(\vp)  - \chi^Q(\mu_\chi).
\ee
We choose the vectors  $\vec{\mu}_d= (\vec{\mu}_{d, \perp}, \mu_{d, n})$  and $\vec{\mu}_\chi =  (\vec{\mu}_{\chi, \perp}, \mu_{\chi, n})$ for given subtraction points $\mu_d = | \vec{\mu}_d|$ and $\mu_\chi = |\vec{\mu}_\chi|$  to be parallel to the external momentum $\vp$, i.e. we put
\be
 \vec{\mu}_d = \mu_d \hat{\vp},\quad \vec{\mu}_\chi = \mu_\chi \hat{\vp},\quad \hat{\vp} \coloneq \frac{\vp}{|\vp|}.
\ee
For a more general discussion of the renormalization of the Hamiltonian approach in Coulomb gauge see Ref.~\cite{Epple:2007ut}, and Ref.~\cite{Heffner:2012sx} for the renormalization at finite temperature.
\begin{figure}[t]
\centering
\subfloat[]{\label{fig:ghost}\includegraphics[width=0.49\linewidth]{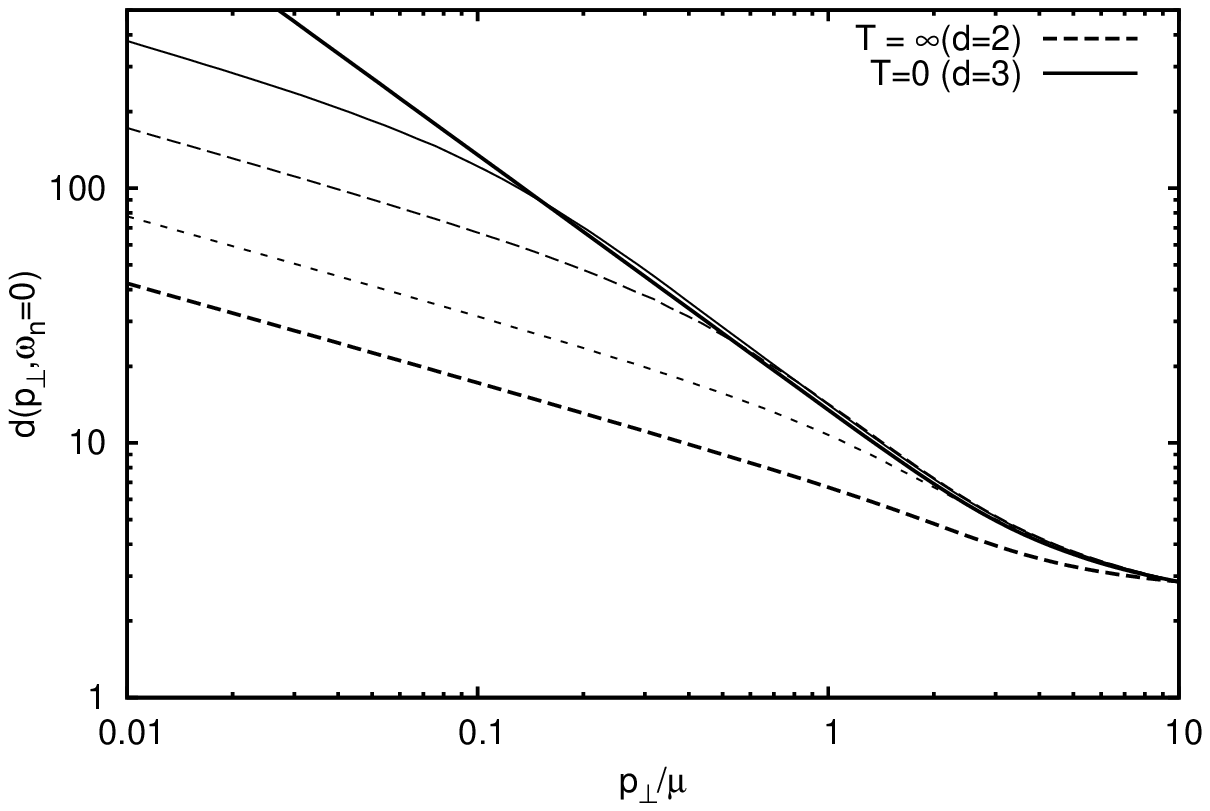}}
\subfloat[]{\label{fig:ghostdiffN} \includegraphics[width=0.49\linewidth]{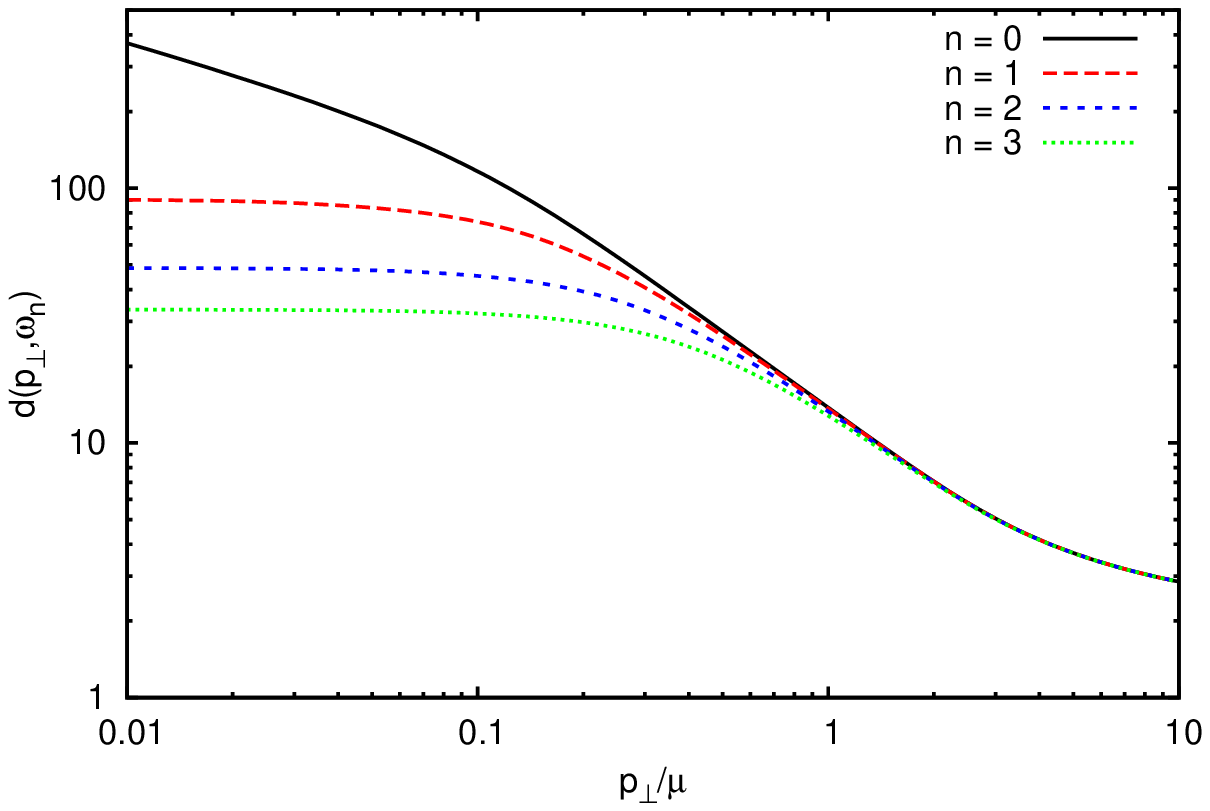}}
\caption{The ghost form factor $d (p_\perp, \omega_n)$ (a) as 
function of the transverse momentum $p_\perp$ for the lowest Matsubara frequency $n = 0$ and for various temperatures decreasing from bottom to top, and (b) at fixed temperature for the first Matsubara frequencies $n = 0,\ldots,3$.} 
\end{figure}
\section{Numerical results}\label{sect:numerical}

Our equations of motion (\ref{81}) and (\ref{eq:ghostren}) represent a set of coupled integral equations with respect to the transverse momentum $\vp_\perp$.
Each of these integral equations labeled by (the index of) the Matsubara frequencies $\omega_n$ is solved using standard methods, see e.q. Ref.~\cite{Epple:2007ut}.
The equations belonging to different Matsubara frequencies are coupled through the loop terms.
For small temperatures many Matsubara frequencies need to be included and the number of integral equations becomes large.
To reduce the numerical expense we solve the integral equations explicitly only up to a maximum Matsubara frequency $\omega_{n_\mathrm{max}}$, while for $\omega_n > \omega_{n_\mathrm{max}}$ the extrapolation
\be
\label{eq:cd:extra}
d(\vp_\perp, \omega_n) = d(\sqrt{\vp_\perp^2 + \omega_n^2 -\omega_{n_\mathrm{max}}^2},\omega_{n_\mathrm{max}})
\ee
is used.
This relation is exact for O($3$) invariant solutions, which depend on $\vp_\perp$ and $\omega_n$ only through the combination $\sqrt{\vp_\perp^2 +\omega_n^2}$.
As expected and as a test calculation with $n=20$ Matsubara frequencies confirms, our solutions become quasi O($3$) invariant for momenta large compared to the temperature, i.e. in particular for $\omega_n \gg T$ or $2 \pi n \gg 1$.
In fact this test calculation shows that the interpolation (\ref{eq:cd:extra}) works very well for $n_\mathrm{max} \geq 4$.
In our numerical calculations we have chosen $n_\mathrm{max}=4$ and included in the sums up to $n= 100$ Matsubara frequencies.

In previous studies \cite{Feuchter:2004mk,Epple:2006hv} at zero temperature $\mu_d =0$ was chosen which allows to explicitly implement the horizon conditon $d^{-1}(0)=0$ into the renormalized ghost DSE.
Depending on the starting function of the iteration one obtains then either the $\beta =1$ or $\beta \approx0.8$ solution.
Here we follow a different road, which actually turns out to be more efficient: We choose $\mu_d$ in the ultraviolet and fine tune $d^{-1}(\mu_d)$ such that the horizon condition $d^{-1}(0)=0$ is fulfilled.
For given $\mu_d \neq 0$ there are two distinct values for $d^{-1}(\mu_d)$ for which the horizon condition is realized, one resulting in the $\beta =1$ solution, the other in the $\beta \approx0.8$ solution.
We consider here only the $\beta =1$ solution, which at $T=0$ yields a linearly rising non-Abelian Coulomb potential.
\begin{figure}[t]
\centering
\subfloat[]{\label{fig:omega_t} \includegraphics[width=.49\linewidth]{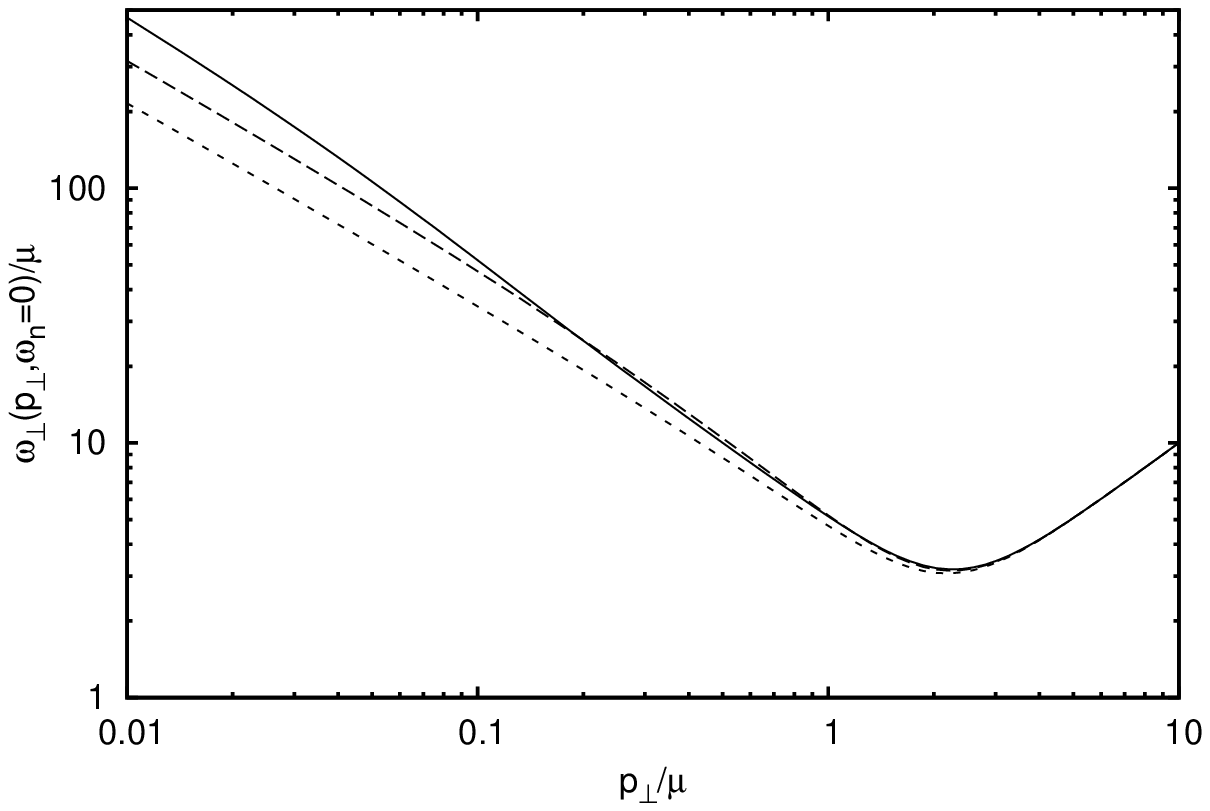}}
\subfloat[]{\label{fig:omega_p} \includegraphics[width=.49\linewidth]{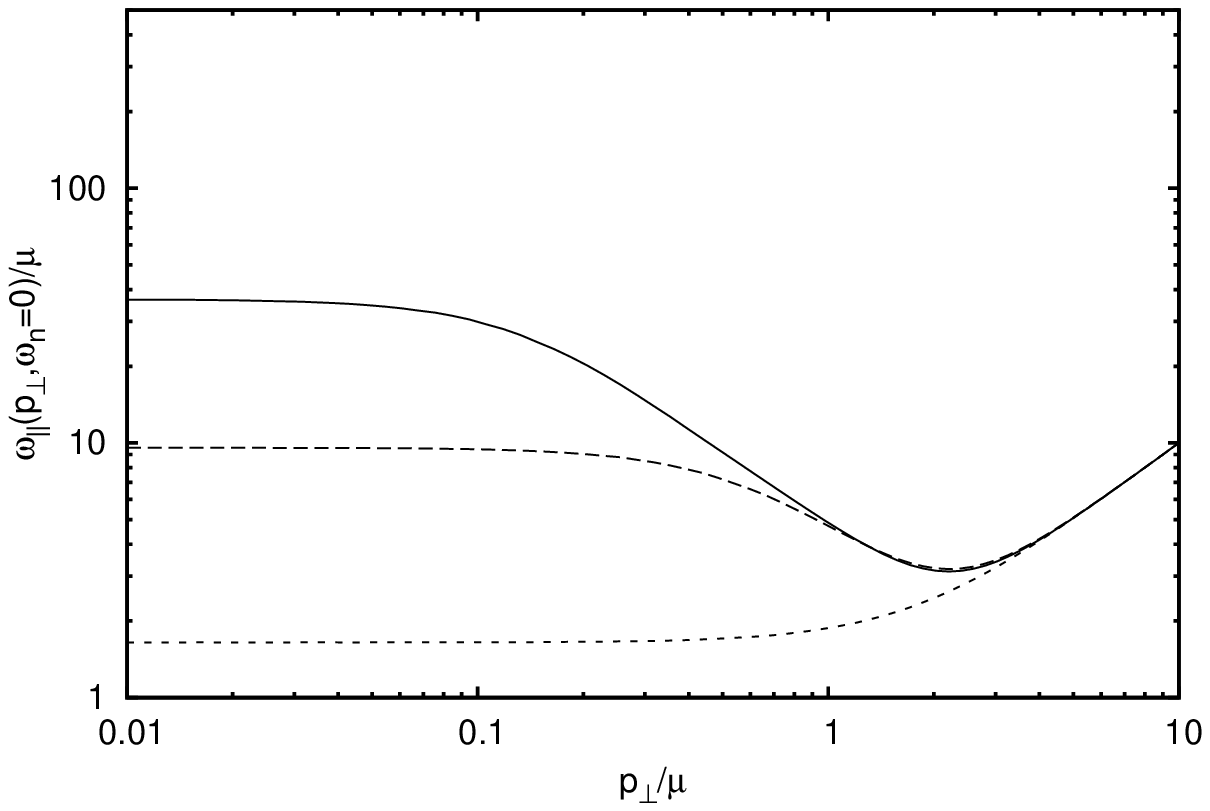}}
\caption{(a) the transversal $\omega_\perp (p_\perp, \omega_n)$ and (b) the longitudinal 
$\omega_{\pa} (p_\perp, \omega_n)$ gluon energies for the lowest Matsubara frequency $n = 0$ as function of the transversal
momentum for various  temperatures decreasing from bottom to top.} \label{fig:omega}
\end{figure}

Fig.~\ref{fig:ghost} shows the ghost form factor for the Matsubara frequency $n = 0$ as function of the ``transverse'' momentum
$p_\perp = | \vp_\perp|$ and for various temperatures.
In the high-temperature limit $d (p_\perp, \omega_n = 0)$ in fact becomes the $T=0$ solution of $d=2$ spatial dimensions and acquires in particular the IR exponent (\ref{527-13-4}) of that solution.
As the temperature is lowered in the mid-to-small-momentum regime a second power-law emerges with an 
exponent given by the IR exponent (\ref{522-13-3}) $\beta =1$ of the $T = 0$ solution of $d = 3$. This power-law extends to smaller and smaller momenta as 
the temperature is lowered, while the power law of the ($T=0$, $d=2$) solution stills persists in the deep IR for any finite temperature and disappears only in the zero-temperature limit
$(L \to \infty)$.

\begin{figure}[t]
\centering
\subfloat[]{\label{fig:sommeN_t} \includegraphics[width=.49\linewidth]{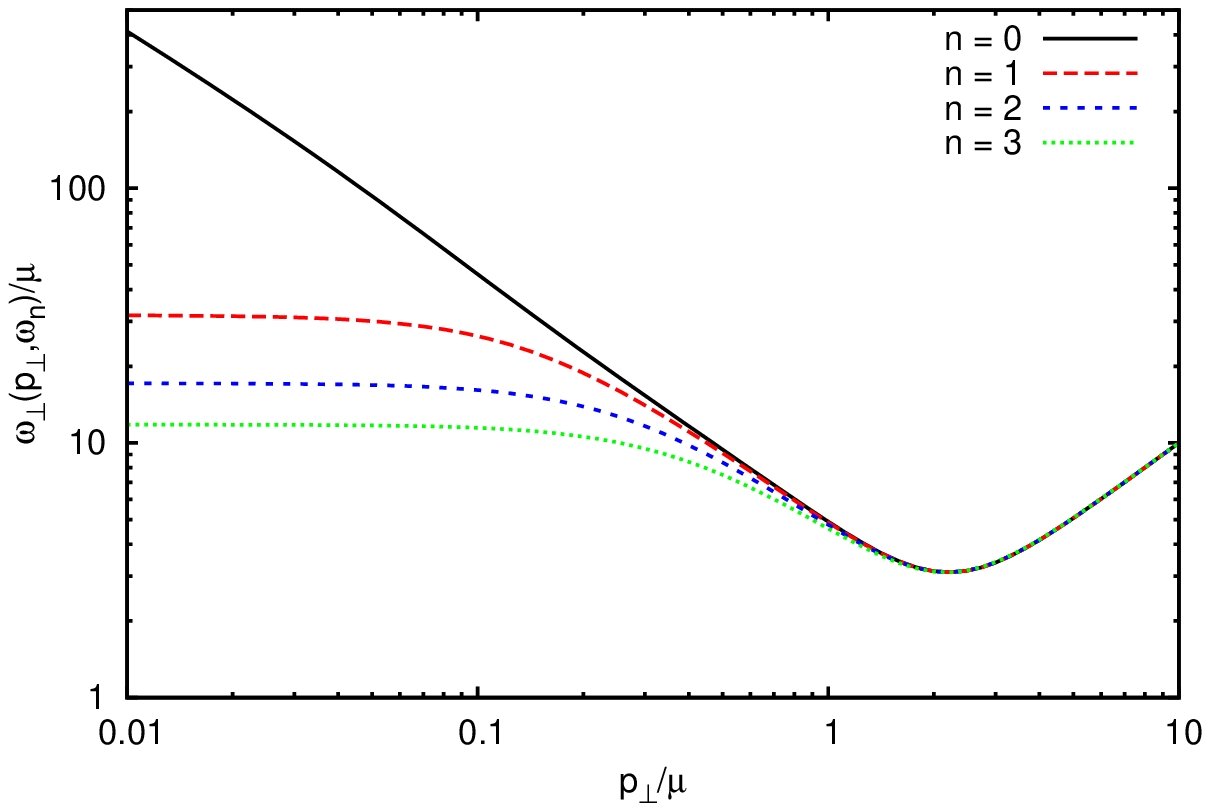}}
\subfloat[]{\label{fig:sommeN_p }\includegraphics[width=.49\linewidth]{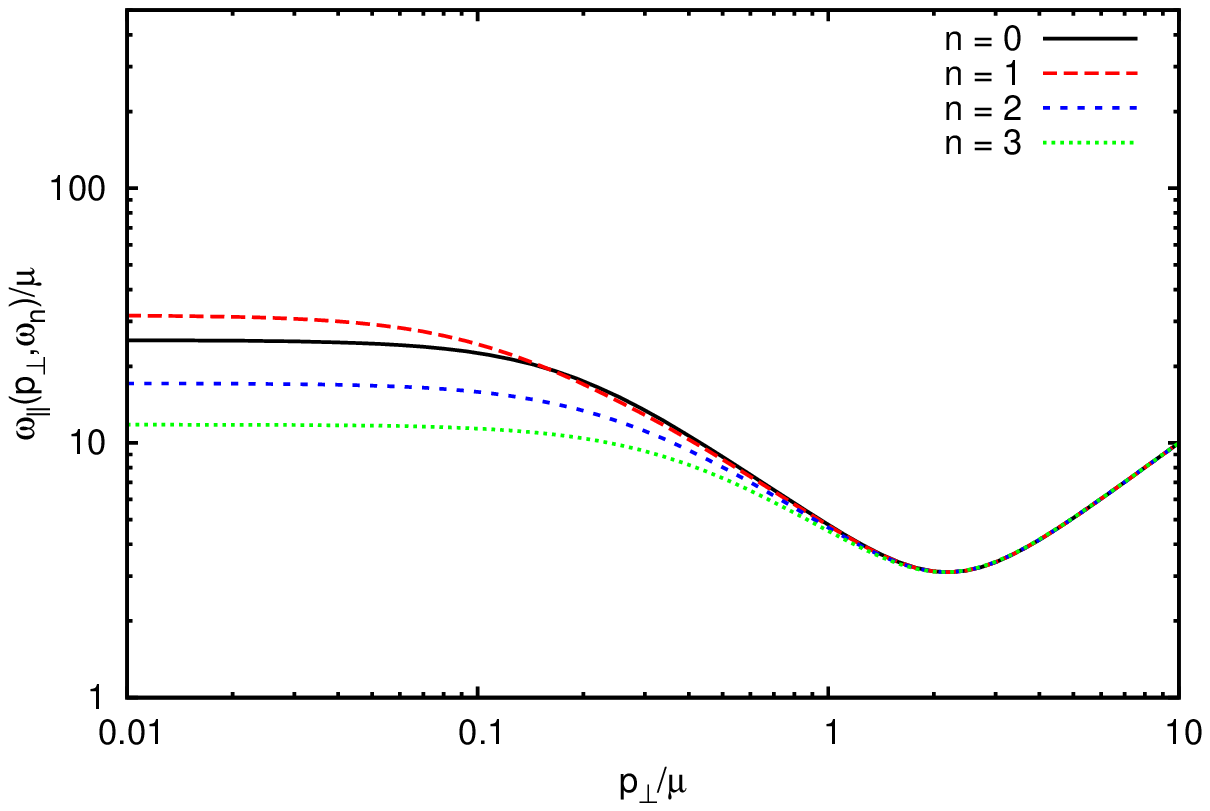}}
\caption{(a) the transversal $\omega_\perp (p_\perp, \omega_n)$  and (b) the longitudinal 
$\omega_{\pa} (p_\perp, \omega_n)$ gluon energies at finite temperature as function of the transversal
momentum and for the first Matsubara frequencies $n = 0,\ldots,3$.} \label{fig:sommeN}
\end{figure}

An analogous temperature behavior is found for the transversal and longitudinal gluon energies shown in Fig.~\ref{fig:omega}. At 
high-temperatures $\omega_\perp (p_\perp, \omega_n = 0)$ approaches the ($T = 0$, $d = 2$)-solution, while $\omega_{\pa} (p_\perp, \omega_n = 0)$ is IR finite. As 
the temperature is lowered, in the mid-to-small momentum regime for both $\omega_{\perp} (p_\perp, \omega_n = 0)$ and $\omega_{\pa} (p_\perp, \omega_n = 0)$ a power-law behavior $\omega_Q (p_\perp, \omega_n = 0)\sim p_\perp^\alpha$ with the $d = 3$ ($T = 0$) IR exponent $\alpha = 1$ emerges.
However, in the deep IR $\omega_Q (p_\perp, \omega_n = 0)$ behaves still as in the high-temperature limit, i.e. $\omega_{\perp} (p_\perp, \omega_n = 0)$  shows the power law with the $d = 2$ IR exponent $\alpha =0.8$ while $\omega_{\pa} (p_\perp, \omega_n = 0)$ goes to a plateau for $p_\perp \to 0$.
With decreasing temperatures the $d=3$-power-law covers an increasing portion of the IR-momentum regime and for $T \to 0$
both $\omega_\perp (p_\perp, \omega_n = 0)$ and $\omega_{\pa} (p_\perp, \omega_n = 0)$ merge to
 the O$(3)$ invariant (IR-diverging) ($T = 0$, $d = 3$)-solution, in accord with our findings for the zero-temperature limit given in Sect.~\ref{sectionIII}.
Finally Fig.~\ref{fig:sommeN} shows the gluon energies $\omega_\perp$ and $\omega_\pa$ at finite temperatures and for the first few Matsubara frequencies $n = 0,\ldots, 3$.
For $n >0$ both gluon energies as well as the ghost form factor [Fig.~(\ref{fig:ghostdiffN})] are infrared finite.

\section{Conclusions}\label{sect:concl}
In this paper we have studied finite-temperature Yang-Mills theory by compactifying a spatial dimension to a circle  $\sphere^1 (L)$ and interpreting its circumference $L$ as the inverse temperature.
In this approach the finite-temperature field theory is entirely encoded in its vacuum state calculated on spatial manifold $\RR^2 \times \sphere^1 (L)$.
The vacuum wave functional on the $\RR^2 \times \sphere^1 (L)$ was determined by the variational principle using a Gaussian type trial state which was generalized to have different variational kernels for the gauge field degrees of freedom parallel and orthogonal to the compactified dimension.
The horizon condition was imposed for the zeroth Matsubara frequency.
At high temperatures $T\to \infty$ the self-consistent solutions approach those of the $d=2$ dimensional theory at $T=0$, as expected from dimensional reduction.
As the temperature is lowered the self-consistent solution exhibits in the medium-to-small momentum regime a power-law behavior with the IR exponent of the $(T=0, d=3)$-solutions, while in the deep IR the ($T=0, d=2$)-power-law still persists.
The IR momentum range of the latter decreases with decreasing temperature and vanishes only in the zero-temperature limit.

The present approach to finite-temperature quantum field theory is advantageous compared to the standard approach based on the grand canonical ensemble since it requires no ansatz for the density matrix.
The presently obtained self-consistent variational solution of Yang-Mills theory on $\RR^2 \times \sphere^1 (L)$ are required as input for a fully self-consistent calculation of the Polyakov loop at finite temperature in the Hamiltonian approach.
This will be subject to future work.

\section*{Acknowledgements:}

The authors are grateful to Davide Campagnari and Markus Quandt for a critical reading of the manuscript and useful comments.

This work was supported by Deutsche Forschungsgemeinschaft under contract DFG-Re 856-9/1.
\bibliography{bib}

\begin{thebibliography}{20}%
\makeatletter
\providecommand \@ifxundefined [1]{%
 \@ifx{#1\undefined}
}%
\providecommand \@ifnum [1]{%
 \ifnum #1\expandafter \@firstoftwo
 \else \expandafter \@secondoftwo
 \fi
}%
\providecommand \@ifx [1]{%
 \ifx #1\expandafter \@firstoftwo
 \else \expandafter \@secondoftwo
 \fi
}%
\providecommand \natexlab [1]{#1}%
\providecommand \enquote  [1]{``#1''}%
\providecommand \bibnamefont  [1]{#1}%
\providecommand \bibfnamefont [1]{#1}%
\providecommand \citenamefont [1]{#1}%
\providecommand \href@noop [0]{\@secondoftwo}%
\providecommand \href [0]{\begingroup \@sanitize@url \@href}%
\providecommand \@href[1]{\@@startlink{#1}\@@href}%
\providecommand \@@href[1]{\endgroup#1\@@endlink}%
\providecommand \@sanitize@url [0]{\catcode `\\12\catcode `\$12\catcode
  `\&12\catcode `\#12\catcode `\^12\catcode `\_12\catcode `\%12\relax}%
\providecommand \@@startlink[1]{}%
\providecommand \@@endlink[0]{}%
\providecommand \url  [0]{\begingroup\@sanitize@url \@url }%
\providecommand \@url [1]{\endgroup\@href {#1}{\urlprefix }}%
\providecommand \urlprefix  [0]{URL }%
\providecommand \Eprint [0]{\href }%
\providecommand \doibase [0]{http://dx.doi.org/}%
\providecommand \selectlanguage [0]{\@gobble}%
\providecommand \bibinfo  [0]{\@secondoftwo}%
\providecommand \bibfield  [0]{\@secondoftwo}%
\providecommand \translation [1]{[#1]}%
\providecommand \BibitemOpen [0]{}%
\providecommand \bibitemStop [0]{}%
\providecommand \bibitemNoStop [0]{.\EOS\space}%
\providecommand \EOS [0]{\spacefactor3000\relax}%
\providecommand \BibitemShut  [1]{\csname bibitem#1\endcsname}%
\let\auto@bib@innerbib\@empty
%</preamble>
\bibitem [{\citenamefont {Feuchter}\ and\ \citenamefont
  {Reinhardt}(2004{\natexlab{a}})}]{Feuchter:2004mk}%
  \BibitemOpen
  \bibfield  {author} {\bibinfo {author} {\bibfnamefont {C.}~\bibnamefont
  {Feuchter}}\ and\ \bibinfo {author} {\bibfnamefont {H.}~\bibnamefont
  {Reinhardt}},\ }\href@noop {} {\bibfield  {journal} {\bibinfo  {journal}
  {Phys.Rev.}\ }\textbf {\bibinfo {volume} {D70}},\ \bibinfo {pages} {105021}
  (\bibinfo {year} {2004}{\natexlab{a}})}\BibitemShut {NoStop}%
\bibitem [{\citenamefont {Feuchter}\ and\ \citenamefont
  {Reinhardt}(2004{\natexlab{b}})}]{Feuchter:2004gb}%
  \BibitemOpen
  \bibfield  {author} {\bibinfo {author} {\bibfnamefont {C.}~\bibnamefont
  {Feuchter}}\ and\ \bibinfo {author} {\bibfnamefont {H.}~\bibnamefont
  {Reinhardt}},\ }\href@noop {} {\  (\bibinfo {year} {2004}{\natexlab{b}})},\
  \Eprint {http://arxiv.org/abs/hep-th/0402106} {arXiv:hep-th/0402106 [hep-th]}
  \BibitemShut {NoStop}%
%%CITATION = HEP-TH/0402106;%%
\bibitem [{\citenamefont {Reinhardt}\ and\ \citenamefont
  {Feuchter}(2005)}]{Reinhardt:2004mm}%
  \BibitemOpen
  \bibfield  {author} {\bibinfo {author} {\bibfnamefont {H.}~\bibnamefont
  {Reinhardt}}\ and\ \bibinfo {author} {\bibfnamefont {C.}~\bibnamefont
  {Feuchter}},\ }\href@noop {} {\bibfield  {journal} {\bibinfo  {journal}
  {Phys.Rev.}\ }\textbf {\bibinfo {volume} {D71}},\ \bibinfo {pages} {105002}
  (\bibinfo {year} {2005})}\BibitemShut {NoStop}%
\bibitem [{\citenamefont {Schutte}(1985)}]{Schutte:1985sd}%
  \BibitemOpen
  \bibfield  {author} {\bibinfo {author} {\bibfnamefont {D.}~\bibnamefont
  {Schutte}},\ }\href {\doibase 10.1103/PhysRevD.31.810} {\bibfield  {journal}
  {\bibinfo  {journal} {Phys.Rev.}\ }\textbf {\bibinfo {volume} {D31}},\
  \bibinfo {pages} {810} (\bibinfo {year} {1985})}\BibitemShut {NoStop}%
%%CITATION = PHRVA,D31,810;%%
\bibitem [{\citenamefont {Szczepaniak}\ and\ \citenamefont
  {Swanson}(2002)}]{Szczepaniak:2001rg}%
  \BibitemOpen
  \bibfield  {author} {\bibinfo {author} {\bibfnamefont {A.~P.}\ \bibnamefont
  {Szczepaniak}}\ and\ \bibinfo {author} {\bibfnamefont {E.~S.}\ \bibnamefont
  {Swanson}},\ }\href@noop {} {\bibfield  {journal} {\bibinfo  {journal}
  {Phys.Rev.}\ }\textbf {\bibinfo {volume} {D65}},\ \bibinfo {pages} {025012}
  (\bibinfo {year} {2002})}\BibitemShut {NoStop}%
\bibitem [{\citenamefont {Greensite}\ \emph {et~al.}(2011)\citenamefont
  {Greensite}, \citenamefont {Matevosyan}, \citenamefont {Olejnik},
  \citenamefont {Quandt}, \citenamefont {Reinhardt} \emph
  {et~al.}}]{Greensite:2011pj}%
  \BibitemOpen
  \bibfield  {author} {\bibinfo {author} {\bibfnamefont {J.}~\bibnamefont
  {Greensite}}, \bibinfo {author} {\bibfnamefont {H.}~\bibnamefont
  {Matevosyan}}, \bibinfo {author} {\bibfnamefont {S.}~\bibnamefont {Olejnik}},
  \bibinfo {author} {\bibfnamefont {M.}~\bibnamefont {Quandt}}, \bibinfo
  {author} {\bibfnamefont {H.}~\bibnamefont {Reinhardt}},  \emph {et~al.},\
  }\href {\doibase 10.1103/PhysRevD.83.114509} {\bibfield  {journal} {\bibinfo
  {journal} {Phys.Rev.}\ }\textbf {\bibinfo {volume} {D83}},\ \bibinfo {pages}
  {114509} (\bibinfo {year} {2011})},\ \Eprint {http://arxiv.org/abs/1102.3941}
  {arXiv:1102.3941 [hep-lat]} \BibitemShut {NoStop}%
%%CITATION = ARXIV:1102.3941;%%
\bibitem [{\citenamefont {Quandt}\ \emph {et~al.}(2014)\citenamefont {Quandt},
  \citenamefont {Reinhardt},\ and\ \citenamefont {Heffner}}]{Quandt:2013wna}%
  \BibitemOpen
  \bibfield  {author} {\bibinfo {author} {\bibfnamefont {M.}~\bibnamefont
  {Quandt}}, \bibinfo {author} {\bibfnamefont {H.}~\bibnamefont {Reinhardt}}, \
  and\ \bibinfo {author} {\bibfnamefont {J.}~\bibnamefont {Heffner}},\
  }\href@noop {} {\bibfield  {journal} {\bibinfo  {journal} {Phys.Rev.}\
  }\textbf {\bibinfo {volume} {D89}},\ \bibinfo {pages} {065037} (\bibinfo
  {year} {2014})}\BibitemShut {NoStop}%
\bibitem [{\citenamefont {Epple}\ \emph {et~al.}(2007)\citenamefont {Epple},
  \citenamefont {Reinhardt},\ and\ \citenamefont
  {Schleifenbaum}}]{Epple:2006hv}%
  \BibitemOpen
  \bibfield  {author} {\bibinfo {author} {\bibfnamefont {D.}~\bibnamefont
  {Epple}}, \bibinfo {author} {\bibfnamefont {H.}~\bibnamefont {Reinhardt}}, \
  and\ \bibinfo {author} {\bibfnamefont {W.}~\bibnamefont {Schleifenbaum}},\
  }\href@noop {} {\bibfield  {journal} {\bibinfo  {journal} {Phys.Rev.}\
  }\textbf {\bibinfo {volume} {D75}},\ \bibinfo {pages} {045011} (\bibinfo
  {year} {2007})}\BibitemShut {NoStop}%
\bibitem [{\citenamefont {Reinhardt}\ and\ \citenamefont
  {Epple}(2007)}]{Reinhardt:2007wh}%
  \BibitemOpen
  \bibfield  {author} {\bibinfo {author} {\bibfnamefont {H.}~\bibnamefont
  {Reinhardt}}\ and\ \bibinfo {author} {\bibfnamefont {D.}~\bibnamefont
  {Epple}},\ }\href@noop {} {\bibfield  {journal} {\bibinfo  {journal}
  {Phys.Rev.}\ }\textbf {\bibinfo {volume} {D76}},\ \bibinfo {pages} {065015}
  (\bibinfo {year} {2007})}\BibitemShut {NoStop}%
\bibitem [{\citenamefont {Reinhardt}(2008)}]{Reinhardt:2008ek}%
  \BibitemOpen
  \bibfield  {author} {\bibinfo {author} {\bibfnamefont {H.}~\bibnamefont
  {Reinhardt}},\ }\href@noop {} {\bibfield  {journal} {\bibinfo  {journal}
  {Phys.Rev.Lett.}\ }\textbf {\bibinfo {volume} {101}},\ \bibinfo {pages}
  {061602} (\bibinfo {year} {2008})}\BibitemShut {NoStop}%
\bibitem [{\citenamefont {Zwanziger}(1998)}]{Zwanziger:1998ez}%
  \BibitemOpen
  \bibfield  {author} {\bibinfo {author} {\bibfnamefont {D.}~\bibnamefont
  {Zwanziger}},\ }\href {\doibase 10.1016/S0550-3213(98)00031-5} {\bibfield
  {journal} {\bibinfo  {journal} {Nucl.Phys.}\ }\textbf {\bibinfo {volume}
  {B518}},\ \bibinfo {pages} {237} (\bibinfo {year} {1998})}\BibitemShut
  {NoStop}%
%%CITATION = NUPHA,B518,237;%%
\bibitem [{\citenamefont {Reinhardt}\ \emph {et~al.}(2011)\citenamefont
  {Reinhardt}, \citenamefont {Campagnari},\ and\ \citenamefont
  {Szczepaniak}}]{Reinhardt:2011hq}%
  \BibitemOpen
  \bibfield  {author} {\bibinfo {author} {\bibfnamefont {H.}~\bibnamefont
  {Reinhardt}}, \bibinfo {author} {\bibfnamefont {D.}~\bibnamefont
  {Campagnari}}, \ and\ \bibinfo {author} {\bibfnamefont {A.}~\bibnamefont
  {Szczepaniak}},\ }\href {\doibase 10.1103/PhysRevD.84.045006} {\bibfield
  {journal} {\bibinfo  {journal} {Phys.Rev.}\ }\textbf {\bibinfo {volume}
  {D84}},\ \bibinfo {pages} {045006} (\bibinfo {year} {2011})},\ \Eprint
  {http://arxiv.org/abs/1107.3389} {arXiv:1107.3389 [hep-th]} \BibitemShut
  {NoStop}%
%%CITATION = ARXIV:1107.3389;%%
\bibitem [{\citenamefont {Heffner}\ \emph {et~al.}(2012)\citenamefont
  {Heffner}, \citenamefont {Reinhardt},\ and\ \citenamefont
  {Campagnari}}]{Heffner:2012sx}%
  \BibitemOpen
  \bibfield  {author} {\bibinfo {author} {\bibfnamefont {J.}~\bibnamefont
  {Heffner}}, \bibinfo {author} {\bibfnamefont {H.}~\bibnamefont {Reinhardt}},
  \ and\ \bibinfo {author} {\bibfnamefont {D.~R.}\ \bibnamefont {Campagnari}},\
  }\href@noop {} {\bibfield  {journal} {\bibinfo  {journal} {Phys.Rev.}\
  }\textbf {\bibinfo {volume} {D85}},\ \bibinfo {pages} {125029} (\bibinfo
  {year} {2012})}\BibitemShut {NoStop}%
\bibitem [{\citenamefont {Reinhardt}\ and\ \citenamefont
  {Heffner}(2012)}]{Reinhardt:2012qe}%
  \BibitemOpen
  \bibfield  {author} {\bibinfo {author} {\bibfnamefont {H.}~\bibnamefont
  {Reinhardt}}\ and\ \bibinfo {author} {\bibfnamefont {J.}~\bibnamefont
  {Heffner}},\ }\href@noop {} {\bibfield  {journal} {\bibinfo  {journal}
  {Phys.Lett.}\ }\textbf {\bibinfo {volume} {B718}},\ \bibinfo {pages} {672}
  (\bibinfo {year} {2012})}\BibitemShut {NoStop}%
\bibitem [{\citenamefont {Reinhardt}\ and\ \citenamefont
  {Heffner}(2013)}]{Reinhardt:2013iia}%
  \BibitemOpen
  \bibfield  {author} {\bibinfo {author} {\bibfnamefont {H.}~\bibnamefont
  {Reinhardt}}\ and\ \bibinfo {author} {\bibfnamefont {J.}~\bibnamefont
  {Heffner}},\ }\href@noop {} {\bibfield  {journal} {\bibinfo  {journal}
  {Phys.Rev.}\ }\textbf {\bibinfo {volume} {D88}},\ \bibinfo {pages} {045024}
  (\bibinfo {year} {2013})}\BibitemShut {NoStop}%
\bibitem [{\citenamefont {Christ}\ and\ \citenamefont
  {Lee}(1980)}]{Christ:1980ku}%
  \BibitemOpen
  \bibfield  {author} {\bibinfo {author} {\bibfnamefont {N.}~\bibnamefont
  {Christ}}\ and\ \bibinfo {author} {\bibfnamefont {T.}~\bibnamefont {Lee}},\
  }\href {\doibase 10.1103/PhysRevD.22.939} {\bibfield  {journal} {\bibinfo
  {journal} {Phys.Rev.}\ }\textbf {\bibinfo {volume} {D22}},\ \bibinfo {pages}
  {939} (\bibinfo {year} {1980})}\BibitemShut {NoStop}%
%%CITATION = PHRVA,D22,939;%%
\bibitem [{\citenamefont {Lerche}\ and\ \citenamefont {von
  Smekal}(2002)}]{Lerche:2002ep}%
  \BibitemOpen
  \bibfield  {author} {\bibinfo {author} {\bibfnamefont {C.}~\bibnamefont
  {Lerche}}\ and\ \bibinfo {author} {\bibfnamefont {L.}~\bibnamefont {von
  Smekal}},\ }\href {\doibase 10.1103/PhysRevD.65.125006} {\bibfield  {journal}
  {\bibinfo  {journal} {Phys.Rev.}\ }\textbf {\bibinfo {volume} {D65}},\
  \bibinfo {pages} {125006} (\bibinfo {year} {2002})},\ \Eprint
  {http://arxiv.org/abs/hep-ph/0202194} {arXiv:hep-ph/0202194 [hep-ph]}
  \BibitemShut {NoStop}%
%%CITATION = HEP-PH/0202194;%%
\bibitem [{\citenamefont {Schleifenbaum}\ \emph {et~al.}(2006)\citenamefont
  {Schleifenbaum}, \citenamefont {Leder},\ and\ \citenamefont
  {Reinhardt}}]{Schleifenbaum:2006bq}%
  \BibitemOpen
  \bibfield  {author} {\bibinfo {author} {\bibfnamefont {W.}~\bibnamefont
  {Schleifenbaum}}, \bibinfo {author} {\bibfnamefont {M.}~\bibnamefont
  {Leder}}, \ and\ \bibinfo {author} {\bibfnamefont {H.}~\bibnamefont
  {Reinhardt}},\ }\href@noop {} {\bibfield  {journal} {\bibinfo  {journal}
  {Phys.Rev.}\ }\textbf {\bibinfo {volume} {D73}},\ \bibinfo {pages} {125019}
  (\bibinfo {year} {2006})}\BibitemShut {NoStop}%
\bibitem [{\citenamefont {Feuchter}\ and\ \citenamefont
  {Reinhardt}(2008)}]{Feuchter:2007mq}%
  \BibitemOpen
  \bibfield  {author} {\bibinfo {author} {\bibfnamefont {C.}~\bibnamefont
  {Feuchter}}\ and\ \bibinfo {author} {\bibfnamefont {H.}~\bibnamefont
  {Reinhardt}},\ }\href {\doibase 10.1103/PhysRevD.77.085023} {\bibfield
  {journal} {\bibinfo  {journal} {Phys.Rev.}\ }\textbf {\bibinfo {volume}
  {D77}},\ \bibinfo {pages} {085023} (\bibinfo {year} {2008})},\ \Eprint
  {http://arxiv.org/abs/0711.2452} {arXiv:0711.2452 [hep-th]} \BibitemShut
  {NoStop}%
%%CITATION = ARXIV:0711.2452;%%
\bibitem [{\citenamefont {Epple}\ \emph {et~al.}(2008)\citenamefont {Epple},
  \citenamefont {Reinhardt}, \citenamefont {Schleifenbaum},\ and\ \citenamefont
  {Szczepaniak}}]{Epple:2007ut}%
  \BibitemOpen
  \bibfield  {author} {\bibinfo {author} {\bibfnamefont {D.}~\bibnamefont
  {Epple}}, \bibinfo {author} {\bibfnamefont {H.}~\bibnamefont {Reinhardt}},
  \bibinfo {author} {\bibfnamefont {W.}~\bibnamefont {Schleifenbaum}}, \ and\
  \bibinfo {author} {\bibfnamefont {A.}~\bibnamefont {Szczepaniak}},\ }\href
  {\doibase 10.1103/PhysRevD.77.085007} {\bibfield  {journal} {\bibinfo
  {journal} {Phys.Rev.}\ }\textbf {\bibinfo {volume} {D77}},\ \bibinfo {pages}
  {085007} (\bibinfo {year} {2008})},\ \Eprint {http://arxiv.org/abs/0712.3694}
  {arXiv:0712.3694 [hep-th]} \BibitemShut {NoStop}%
%%CITATION = ARXIV:0712.3694;%%
\end{thebibliography}%

\end{document}